\documentclass[runningheads]{llncs}

\usepackage[T1]{fontenc}
\usepackage{graphicx}
\usepackage{amsmath}
\usepackage{amssymb}
\usepackage{url}

\usepackage{algorithm}
\usepackage{algorithmicx}
\usepackage{algcompatible}

\usepackage{subcaption}

\usepackage{color}
\usepackage{comment}

\definecolor{darkgreen}{rgb}{0.0, 0.75, 0}

\begin{document}

%Konputagailuen Arkitektura eta Teknologia Saila,\\ 

\title{Acceleration of Parallel Tempering for Markov Chain Monte Carlo methods}
\author{Aingeru Ramos\inst{1,2}\orcidID{0009-0006-9585-7614} \and
Jose A. Pascual\inst{2}\orcidID{0000-0001-5355-6537} \and
Javier Navaridas\inst{2}\orcidID{0000-0001-7272-6597} \and 
Ivan Coluzza\inst{3}\orcidID{0000-0001-7728-6033}}

\authorrunning{A. Ramos et al.}
\institute{
Basque Center on Material, Applications and Nanostructures (BCMaterials)\\
Martina Casiano building. EHU Technology Park. 48940 Leioa\\
\email{aingeru.ramos@ehu.eus}\\
\and
Dept. of Computer Architecture and Technology, University of the Basque Country\\
Manuel Lardizabal Pasealekua 1, 20018 Donostia\\
\email{\{joseantonio.pascual, javier.navaridas\}@ehu.eus}\\
\and
Rice Univeristy\\
6100 Main St, Houston, TX 77005, US\\
\email{ic33@rice.edu}\\
}

\maketitle

\begin{abstract}
Markov Chain Monte Carlo methods are algorithms used to sample probability distributions, commonly used to sample the Boltzmann distribution of physical/chemical models (e.g., protein folding, Ising model, etc.). This allows us to study their properties by sampling the most probable states of those systems. However, the sampling capabilities of these methods are not sufficiently accurate when handling complex configuration spaces. This has resulted in the development of new techniques that improve sampling accuracy, usually at the expense of increasing the computational cost. One of such techniques is Parallel Tempering which improves accuracy by running several replicas which periodically exchange their states. Computationally, this imposes a significant slow-down, which can be counteracted by means of parallelization. These schemes enable MCMC/PT techniques to be run more effectively and allow larger models to be studied. In this work, we present a parallel implementation of Metropolis-Hastings with Parallel Tempering, using OpenMP and CUDA for the parallelization in modern CPUs and GPUs, respectively. The results show a maximum speed-up of 52x using OpenMP with 48 cores, and of 986x speed-up with the CUDA version. Furthermore, the results serve as a basic benchmark to compare a future quantum implementation of the same algorithm.

\end{abstract}

\section{Introduction}
\label{sec:introduction}

The study of physical systems requires resolving equations, derived from the natural laws that govern those systems; a problem that many times is intractable due to {\bf (i)} the nonexistence of a close solution of the equations that describe the system or {\bf (ii)} the complexity of the equations prevents to tackle the problem manually~\cite{b1}. Because of that, the use of computational methods has become fundamental in physics, and in general, in all the fields of science. The use of these kinds of methods can be traced back to the XX century and can be seen as the birth of {\bf computational physics} and its most fundamental tool: {\bf the simulation of physical models}.

It is important to distinguish between {\it physical} simulation and {\it numerical} simulation. {\bf Physical simulation}, the most intuitive of both, consists of the creation of a virtual version (a computational model) of the system to {\bf emulate the behavior} of the real system, following the laws of nature that influence it~\cite{b2}. Due to the complexity of these systems, scientist tend to rely on simplified models to save computational resources and obtain results more rapidly at the expense of precision in the simulation. This type of simulation is widely used in the field of material science and biology for the analysis of the properties of materials and the evolution of protein systems~\cite{b3,b4,b5,b6}. 

In contrast, {\bf numerical simulations} do not emulate the system under study, instead, they approximate the solution of the problem using numerical and statistical methods~\cite{b7}\cite{b8}, and obtain macroscopic characteristics of the system, such as pressure, temperature, density, or other relevant properties for understanding of the behavior of the system.
In this context, \textbf{Markov Chain Monte Carlo methods (MCMC)} are widely used. Its first development was the {\bf Metropolis-Hastings algorithm (MH)}, created from the works of Metropolis~\cite{b9} and Hastings~\cite{b10}. MH defines the general scheme of all MCMC processes. 
%Indeed, it is common to refer to MH as MCMC algorithm in the literature. 
These algorithms, capable of sampling any given probability distribution, are used as methods of numerical simulations in physical-chemical systems together with the Boltzmann distribution that describes the probability of a system being in any given particular state. Due to that, sampling the Boltzmann distribution through MCMC methods allows the exploration of the state space of the system, thus, obtaining a {\bf sample of the most probable states}. However, in real systems, where the state space is immense and states can be multidimensional, the effectiveness of these methods decreases, returning as a result a {\bf non-representative} sample of the objective distribution.

To improve the effectiveness of these methods, the Parallel Tempering (PT) technique and its derivatives~\cite{b11,b12}  were developed to enhance state space exploration. These kinds of techniques execute multiple MCMC processes at different {\it temperatures}, allowing the swap of information between them~\cite{b11,b13} and, thus, exploring the state space in a more effective manner. However, the computational cost multiplies due to the increase in {\bf concurrent MCMC processes}. The parallelization of the PT technique has been dealt with mainly by employing parallel architectures to speed up specific phases of the MCMC methods and reduce the execution time of the concurrent processes~\cite{b14}. In these cases, the huge parallelism offered by GPUs, is very commonly taken advantage of. However, in most cases, all the information of the simulation is stored in the memory of the {\it host} system, using the GPU(s) only for computing, which slows down simulations due to the costly data transfers between the host and the accelerator. Furthermore, new variants of the MH algorithm are proposed, such as Gibbs sampling~\cite{b15} which reduces the number of iterations needed to obtain a sample, again, at the expense of increasing the computational cost of every individual MH process. Another important area is the parallelization of the swaps between the MCMC processes inside PT which poses major {\bf synchronization issues} and, in turn, increases the complexity of parallelization.

In this paper, we present a Metropolis-Hastings (MH) algorithm implementation using Parallel Tempering (PT) on parallel architectures. Our approach consists in distributing the MH processes along all available threads, parallelizing the swaps of the processes inside PT technique. The algorithm has been implemented in two paradigms: {\bf (i)}  OpenMP for multiprocessors and multicore; {\bf (ii)} CUDA for GPUs. In the later one, all information is stored in the {\it device} memory to minimize {\it host} memory access latency. To evaluate the performance of the implementations, we have modeled a two-dimensional Ising model. Both OpenMP and CUDA implementations demonstrate {\bf good scalability}, achieving 2 and 3 orders of magnitude speed-up factors, respectively. One important thing to note is that the main objective of this work is not to develop the most efficient implementation of the algorithm, but to {\bf create a benchmark} that considers both performance and convergence times, thus enabling the comparison of this classical algorithm with future work related to the thesis in which this study is framed: the development of a {\bf quantum version of the MCMC algorithm}.

\section{Theoretical Framework}
\label{sec:theoretical_framwork}

This section is dedicated to presenting the key concepts necessary to understand the domain of the work.

\subsection{Markov Chain Monte Carlo}
\label{subsec:mcmc}

Let $ X $ be a random variable following a probability distribution $ f(x) $, and let $ \{X_0, X_1, X_2, ..., X_n\} $ be a random sample of size $ n+1 $, where the frequency of appearance of each possible value of X is proportional to the probability given by $ f $. There exists some distributions with well-known methods to sample those functions. However, these cases are atypical and in the vast majority of cases, the sample must be realized using approximation methods, such as {\bf Monte Carlo methods} that allow  obtaining \textit{independent samples} from any given distribution $ f $. In contrast, MCMC also allow the sampling of any distribution, but in these cases, the samples are \textit{dependent} due to the implicit Markov chain constructed by these kinds of methods.

The {\bf basic scheme} of MH algorithm is the following. Suppose a variable,  $ x $, is initialized with a random value, $ x_{init} $. For each of the $ N $ iterations of the process, three steps are executed: {\bf (i)} a trial, $ x_{trial} $, is generated through a \textit{generator} function, $ g$, and the current value of $ x $, resulting in $ x_{trial} = g(x) $. {\bf (ii)} the probability of $ x_{trial} $ substituting the current value of $ x $ is calculated using the acceptance function, $ A $, described in Eq.~\eqref{eq:acc_f} for the MH algorithm. {\bf (iii)} the current value of $ x $ is substituted by $ x_{trial} $ according to the previous probability and the value of $ x $ is stored in a list. At the end of the $ N $ iterations, the list contains the Markov chain generated by this process---a sample of the distribution.

\begin{equation}
    A(x_{trial}, x) = min\left(
        1, \frac{f(x_{trial})}{f(x)}\right)
    \label{eq:acc_f}
\end{equation}

All the algorithms of the MCMC family follow this scheme with slight modifications to $ g $ and/or $ A $. In order for MCMC processes to generate descriptive samples of the distribution, it is fundamental that these processes satisfy the ergodic property. This property depends on $ g $, $ A $, and $ f $. Specifically, this property implies that the process must guarantee that, with enough number of iterations, the process can visit every value of $ f $; i.e., when $ N \rightarrow \infty $ the process must pass through all possible values of $ f $. Apart from ensuring ergodicity, the main challenges of this type of algorithm are its slow convergence rate, the difficulty in obtaining meaningful samples when the $ f $ is multimodal and/or multidimensional, or when the space to be explored is vast or even intractable. In these cases, the number of iterations required becomes so high that the algorithm becomes unfeasible in its basic form. There are improved versions of the method that optimize the convergence rate, such as the {\bf Parallel Tempering (PT)} technique.

%\begin{figure}[t] 
%    \begin{center} 
%    \includegraphics[width=0.7\linewidth]{figs/ergodic.pdf}
%    \end{center} 
%    \caption{Traces (left-hand side) and distributions (right-hand side) of chains resulting from applying different acceptance functions sampling the $\mathcal{N}$(0,1) distribution.} 
%    \label{fig:accp_cmp} 
%\end{figure} 

\begin{figure} [t]
\centering
    \caption{The figure shows the \textit{flattening} effect of the temperature factor. For (a) the temperature factors have no physical correlation; they are merely illustrative.}
\begin{minipage}{0.495\textwidth}
    \centering
    \subcaption{Normal distribution $\mathcal{N}$(0,1)} \label{fig:pt_Texample}
    \includegraphics[width=\linewidth]{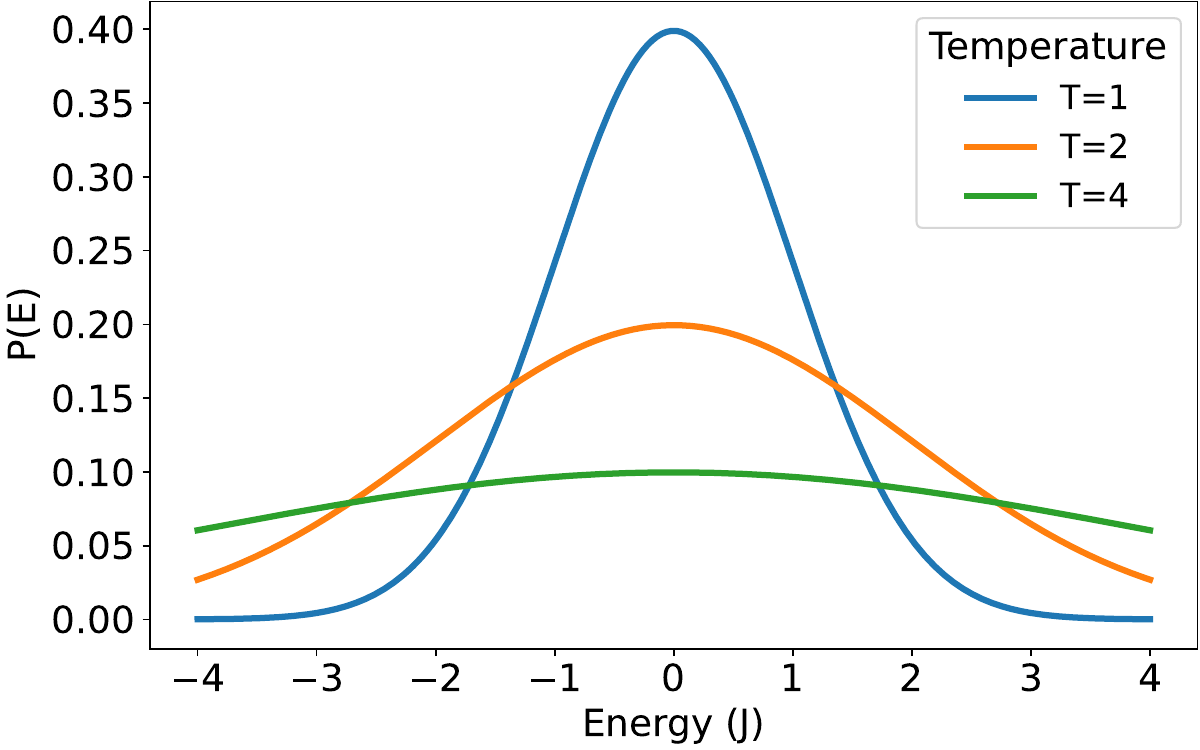}
\end{minipage}
\begin{minipage}{0.495\textwidth}
    \centering
    \subcaption{Boltzmann distribution} \label{fig:boltzmann}
    \includegraphics[width=\linewidth]{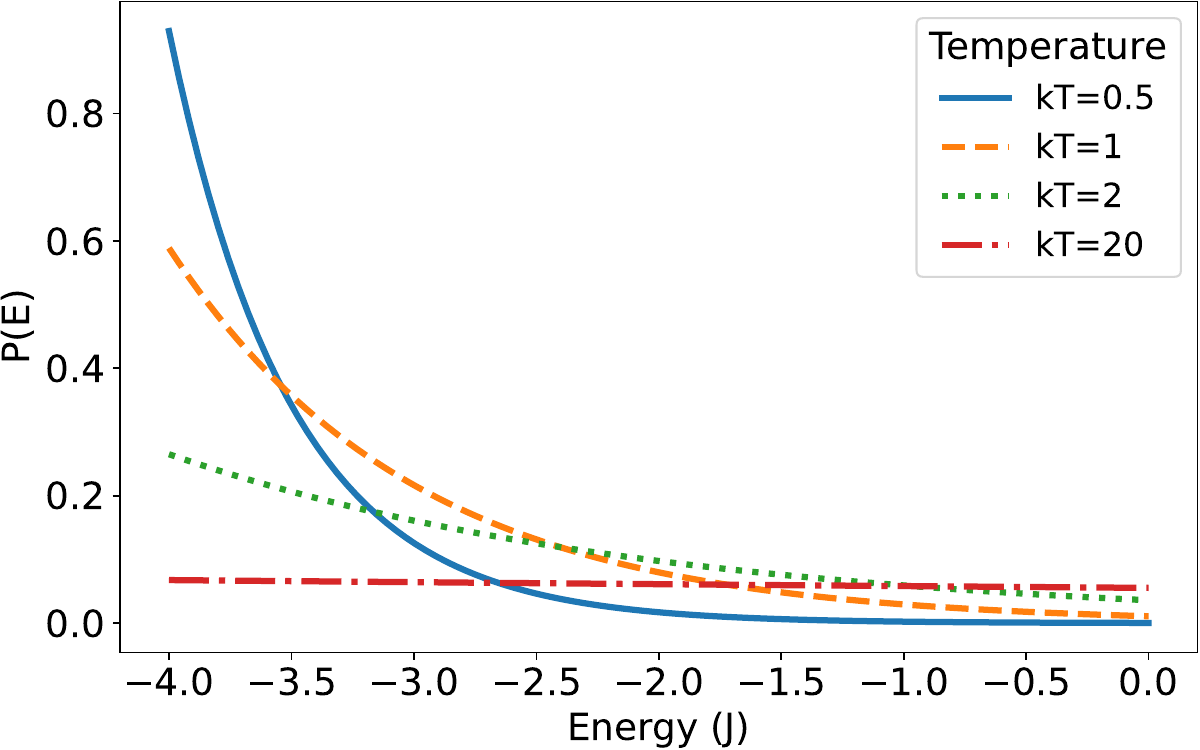}
\end{minipage}
\label{fig:speed_omp_figs}
\end{figure}

Parallel Tempering is a technique that consists in the execution of $ R $ replicas, MCMC independent processes, at different temperatures and allows the interaction between them along the $ N $ iterations of the simulation. The use of different temperatures is inspired by the change of behavior of physical systems at low and high temperatures. Systems fluctuate less at {\bf low temperatures} and, thus, tend to {\bf low energy states}; the ones with the highest probabilities. On the other hand, an {\bf increase in the temperature} causes wider fluctuations, making it easier for systems to reach  {\bf high energy states}, which are less probable. From the statistical point of view, the temperature factor $ T $ acts like a {\bf \textit{flattening} factor} of a distribution. Fig.~\ref{fig:pt_Texample} demonstrates the effect of applying different $ T $ factors to a normal distribution $\mathcal{N}$(0,1).
This results in replicas executed with higher temperatures being capable of exploring the distribution  more effectively. On the other hand, the replicas executed at low temperatures, tend towards the maxima of the distribution. The interaction between replicas consists of {\bf swapping the state} between the high- and low-temperature replicas. This permits replicas with low temperatures to use the extra exploration of the other replicas and reach distribution zones impossible to reach on their own, getting more descriptive samples. The number of iterations between swaps and the probability of performing a swap between two replicas are parameters of the simulations, and there exists no deterministic method to get their optimal values.

\subsection{Ising Model}
\label{subsec:ising}

Statistical mechanics (SM) provides a probabilistic point of view of the evolution of systems, addressing the problem of tracking the evolution of a system composed of a great number of particles, even if the exact state of the system (position and momentum of each particle) is unknown. Under this theoretical framework, the probability of encountering the system in a certain state is determined by the {\bf Boltzmann distribution} \eqref{equ_boltzmann}, a probabilistic distribution that depends on the energy $ E $, and temperature $ T $, of the state. The $ Z $ factor, known as the {\bf partition function}, is necessary to ensure that the summation of the probability of all possible states gives 1 as the result.

\begin{equation}\label{equ_boltzmann}
    P(E_i) = \frac{e^{-E_i/kT}}{Z};\quad Z = \sum_{j}e^{-E_j/kT}
\end{equation}

\noindent
It is worth noticing how the behavior of this distribution varies at different temperatures, as shown in Fig.~\ref{fig:boltzmann}. The systems under low temperatures tend to have configurations with less energy.  On the other hand, higher temperature increases the number and scale of the fluctuations, which enables reaching higher energy configurations.

%\begin{figure}[t]
%    \centering
%    \includegraphics[width=0.7\linewidth]{figs/boltzmann.pdf}
%    \caption{Boltzmann distribution showing the probability of finding the system at a given energy. Note that at excessively high temperatures, the distribution approaches the uniform distribution.}
%    \label{fig:boltzmann}
%\end{figure}

{\bf The Ising model} is well-known and widely used in SM for the analysis and study of materials and their {\bf magnetic behavior}. The model assumes that the magnetism of all materials can be explained by the interactions of spins (variables that have +1 or -1 as possible values) on a 2D or 3D regular lattice. The alignment of neighboring spins contributes to the magnetism, generating a ferromagnetic material, whereas their misalignment  contributes negatively to the magnetism, creating an antiferromagnetic material. The energy formulation for this model is described in Eq.~\eqref{equ_ising}, where $ \sigma $ is the spins configuration of the model, $ B $ is the external magnetic force influencing the material and $ J $ is the interaction force between neighboring spins inside the material, defining ferromagnetic and antiferromagnetic materials.

\begin{equation}
    E(\sigma) = B\sum_{i}{\sigma_i} - J\sum_{\langle i,j \rangle}{\sigma_i \sigma_j}
    \label{equ_ising}
\end{equation}

\noindent
%Analyzing Eq.~\eqref{equ_ising} we can see how $ J $ defines the type of material and its magnetic behavior. 

%Remember that, as shown in Fig.~\ref{fig:boltzmann}, the systems at low temperatures tend to have low energy states. In this case, the model reaches its minimum energy when the spins are aligned and $ J > 0 $, defining a ferromagnetic material. In contrast, if $ J < 0 $, the minimum energy is achieved when the spins are misaligned, describing an antiferromagnetic material. 

Therefore, using \eqref{equ_ising} and the Boltzmann distribution, it is possible to define a probabilistic distribution over all possible energy levels of the model. This, in combination with MH/PT, provides a method to obtain the {\bf most probable configurations} of a material sampling the associated Boltzmann distribution, where the $ x $ values of the chains are $ \sigma $ configurations of the Ising model.

\section{MH/PT Paralelization}
\label{sec:paralelization}

In this paper we present two implementations of the Metropolis-Hastings (MH) algorithm applied in conjunction with Parallel Tempering (PT): the first one in OpenMP for parallel execution on multiprocessors and multicores whereas the second one in CUDA for parallelization on GPUs. The code for both implementations is shared with the community via a git repository\footnote{{\url{https://github.com/AingeruRamos/MCMC_C/}}}.

%\footnote{Anonymized link}%(Removed for anonimization purposes)
%can be accessed using this link\footnote{{\url{https://github.com/AingeruRamos/MCMC_C/}}}.

\begin{figure}[t]
    \centering
    \caption{Representation of the parallelization scheme followed in both implementations. The iterations proceed from left to right (t$_i$). The replicas are represented vertically (R$p$). The boxes represent the iterations to be performed by each replica (in green if they have been executed and in white if they are to be completed). The orange dotted boxes represent {\it Swap iterations} and the orange double-headed arrows represent the organization of exchanges.}
    \includegraphics[width=0.7\linewidth]{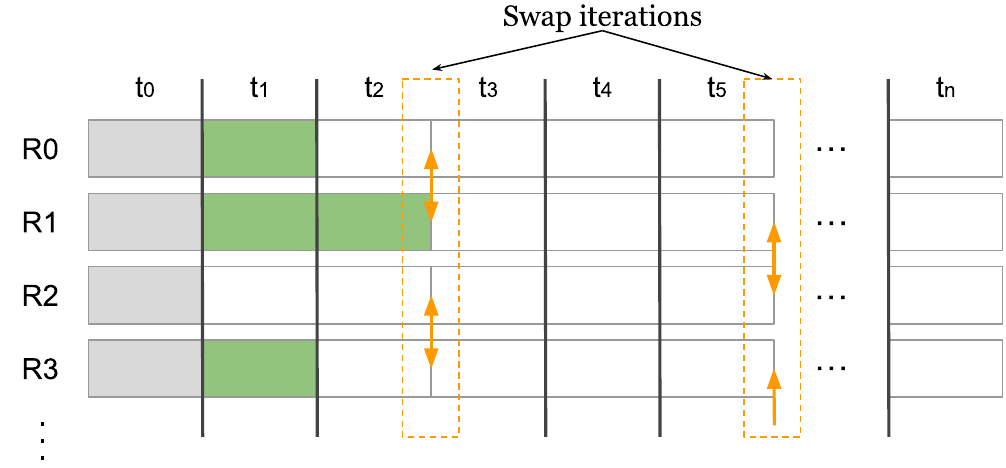}
    \label{fig:parallel_scheme}
\end{figure}

The parallelization scheme is carried out {\bf at replica level}. Let $ R $ be the set of replicas of the simulation. In this paper, we explore the Ising model in the [1.0, 4.0] range of temperature. So, this range will be divided according to the number of replicas $ |R| $ of the simulation; i.e, the temperature assigned to each replica follows the expression $ 1+i\frac{3}{|R|} $; where $ i $ is the index of the replica. Also, each replica runs a unique Ising model instance, but with the same ratio of -1 and +1 spins in the $ \sigma $ configuration and same spin interaction factor J. The replicas are distributed across the available threads, which are responsible for executing the $ n $ iterations of their assigned replicas. Because swapping between replicas requires {\bf synchronizing} replicas, i.e., synchronizing all threads, the computation is {\bf scheduled in intervals} between swap iterations (orange arrows in Fig.~\ref{fig:parallel_scheme}). For instance, in the figure, iterations $ t_3 $, $ t_4 $, and $ t_5 $ conform an interval. Thus, the replicas are executed in parallel by their respective threads until the swap iteration is reached; then execution stops and the replicas interact with each other. This process is repeated until all replicas complete $ n $ iterations.

\begin{comment}
\begin{algorithm}[t]
\caption{Pseudocode of the parallelization scheme of our implementation. The distribution of the replicas is done according to the unique identifier of each thread.}\label{paralllel_scheme}
\begin{algorithmic}[1]
\REQUIRE{Set of replicas $R = \varnothing$ }
\REQUIRE{Number of iterations $n$}

\STATE{$R \gets$ Initialize set of replicas}

\STATE{$i \gets 0$}
\WHILE{$i < n$ }

\STATE{$I\gets$ size of the interval to execute}
\STATE{$\forall r \in R$ \textbf{do} in parallel}
\STATE{$\quad$ execute I iterations of $r$}
\STATE{process the swap between replicas}
\STATE{$iter \gets iter + I$}

\ENDWHILE
\end{algorithmic}
\end{algorithm}
\end{comment}

In each swap iteration, the replicas are paired following two rules: {\bf (i)} each replica can only exchange its state with one of its two neighboring replicas, and {\bf (ii)} in each exchange iteration a replica can only be exchanged once. Thus, the first swap iteration of the simulation allows interaction between replicas $ R_{0}\leftrightarrow R_{1} $, $ R_{2}\leftrightarrow R_{3} $, and so on, as shown in Fig.~\ref{fig:parallel_scheme}. However, in the following swap iteration, the pairing is {\it shifted}, allowing interactions $ R_{1}\leftrightarrow R_{2} $, $ R_{3}\leftrightarrow R_{4} $, etc. (as shown in Fig.~\ref{fig:parallel_scheme}). These pairings alternate to guarantee the propagation of state from one replica to any other replica in a simulation with a sufficient number of iterations. One parameter to be determined is the probability of performing an exchange. 
%Regardless of the expression used to formalize this probability, the expression must maintain the property of ''{\it detailed balance}'' of the process; a necessary property in Markov chains to ensure their convergence, as explained in~\cite{b13}. 
In our particular case, we follow~\cite{b13} and  use a probability based on the energy and temperature differences between the two replicas to determine the exchange probability $ P_{swap}(i,j) = \frac{e^{\Delta \beta_{ij} \Delta E_{ij}}}{1 + e^{\Delta \beta_{ij} \Delta E_{ij}}} $, where $ i $ and $ j $ are two swap candidates and, $ \beta_{ij}$ and $E_{ij} $ the difference of temperature and energy between the candidates, respectively. Finally, the exchange process is also parallelized by distributing the replica pairings among the available threads.

Both versions are divided into two distinct phases: an initialization phase and an execution phase. The initialization phase builds all the data structures necessary for the simulation and calculates the first iteration of all replicas. This is represented as the initial gray rectangles at iteration $ t_0 $ in Fig.~\ref{fig:parallel_scheme}. The execution phase computes the remaining iterations of the replicas, distributing the replicas among the threads as explained above. In the OpenMP version, where the number of threads $H$ is, a priori, much smaller than the number of replicas $ |R| $, each thread is responsible for multiple replicas; namely $ \frac{|R|}{H} $. 
In contrast, in the CUDA version, due to the large number of threads available on GPUs, each replica can be assigned to a single thread, and all of them can be executed at the same time as long as the SM of the device allows it.
Another differentiating point of our CUDA implementation is the management of simulation data: all the simulation information is located inside the GPU, that is, in the {\it device}, thus avoiding transfers to/from the memory of the {\it host}, which would cause delays in the execution of the kernels.

\section{Experimental Results}
\label{sec:exp_results}

As mentioned at the end of the introduction (section \ref{sec:introduction}), the main objective of this work is to develop a benchmark that allows for comparing the classical implementation of the Metropolis-Hastings algorithm with a quantum version still under development. This benchmark focuses on: {\bf (i)} the convergence speed of the replicas and {\bf (ii)} the execution time of the simulation. Special attention is given to studying the convergence speed of the replicas, as quantum computing, leveraging the superposition of qubit states, promises quadratic acceleration in space exploration, which could positively impact the convergence speed. However, it is also necessary to pay attention to the execution time and evaluate the temporal cost of applying such quantum circuits and whether it is temporally profitable, especially considering the current challenges of the technology, such as quantum noise and limitations in the quality of the available hardware.

\subsection{Convergence of the Ising model}
\label{subsec:conver}

\begin{figure}[t]
\centering
    \caption{Both subfigures show information related to the convergence of the simulation replicas. Subfigure (a) shows the percentage of magnetization for each temperature in the Ising model. Subfigure (b) shows the relationship between the number of iterations and the model size. The X-axis represents the size $ L $ of the two-dimensional Ising model, such that $ L \times L $ corresponds to the total number of spins.}
\begin{minipage}{0.495\textwidth}
    \centering
    \subcaption{} \label{fig:ising_phase}
    \includegraphics[width=\linewidth]{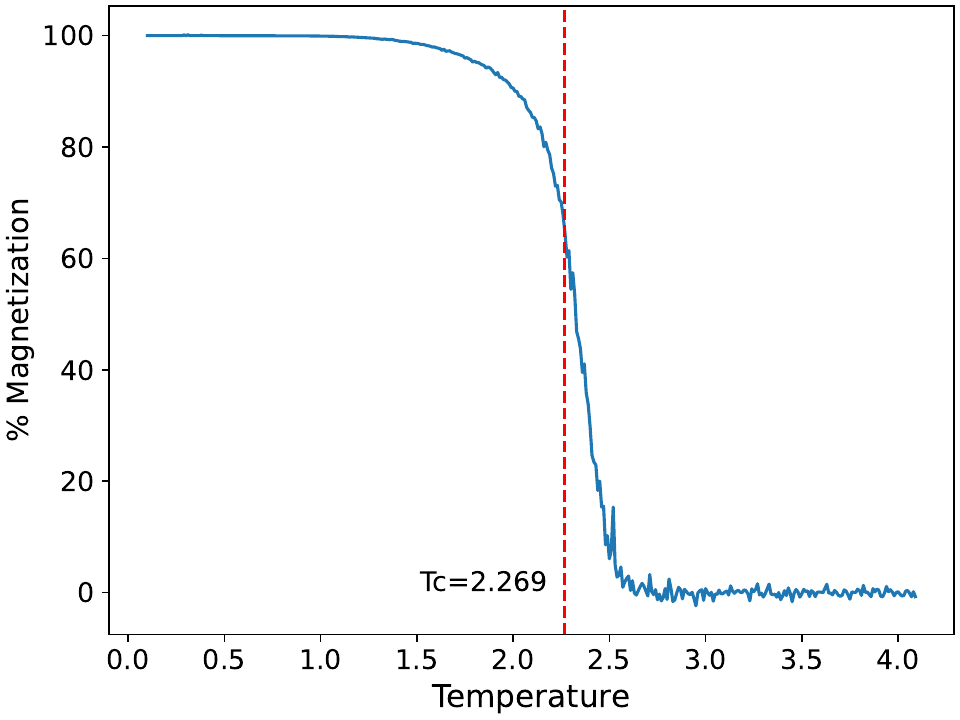}
\end{minipage}
\begin{minipage}{0.495\textwidth}
    \centering
    \subcaption{} \label{fig:conver}
    \includegraphics[width=\linewidth]{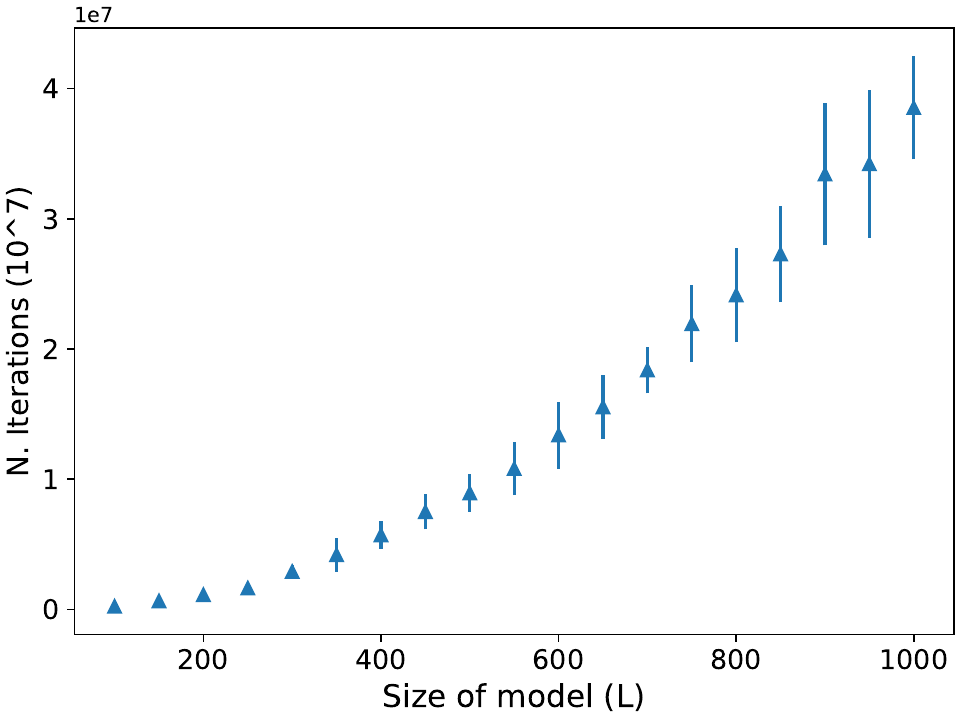}
\end{minipage}
\label{fig:figs}
\end{figure}

As indicated in subsection \ref{subsec:ising}, the Ising model, along with the Boltzmann distribution, is used to study the magnetic properties of materials, allowing for the definition of a probability distribution of the magnetic strength as a function of temperature. Figure \ref{fig:ising_phase} illustrates how the magnetic behavior of the Ising model changes with temperature. The percentage shown in the figure represents the magnetic intensity of the model, expressed as a fraction of the maximum possible value, which occurs when all the spins are fully aligned. It can be observed that, as the Ising model approaches a critical temperature, it undergoes a dramatic change in its magnetic behavior, known as a phase transition, shifting from an ordered state to a disordered one, which causes a significant decrease in magnetic strength. This behavior matches both theoretical predictions and empirical experiments conducted on the model.

Figure \ref{fig:conver} shows the number of iterations required for the replicas to converge to the target distribution as a function of the size of the Ising model used. It is important to note that a two-dimensional Ising model is employed, so when a model is described as having size \(L\), it actually refers to a model with a total number of spins equal to \(L \times L\). Upon analyzing the figure, two points can be observed: (i) as the size of the model increases, the variability in the number of iterations required for convergence also increases, which is due to the stochastic nature of the MCMC algorithm and the greater number of possible states in the system that the Markov chains must navigate to reach convergence; and (ii) a quadratic relationship between the number of iterations and the size of the model. The latter is of great relevance and precisely represents one of the problems that the quantum version of the method aims to address. The quadratic acceleration of quantum circuits could allow the quadratic relationship between the number of iterations and the size of the model to be transformed into a linear relationship, which would imply a drastic reduction in the number of iterations, especially when working with large models.

\subsection{Performance Analysis}
\label{subsec:performance}

To evaluate the performance of our implementation, an Ising model with 90,000 spins distributed on a 300x300 grid was used, with values of $ J = 1 $ and $ B = 0 $, which defines a ferromagnetic material with no external interaction, and constitutes one of the simplest models. The number of iterations per replica was 300,000. Two sets of experiments were carried out: one without swaps and another with swaps. The first is used to measure the degree of parallelization of the implementation, without considering the effect of synchronization between replicas. The second, on the other hand, serves to evaluate this effect when performing swaps, varying the swap frequency with values of 0, 100, 1k, and 10k, and performing a set number of iterations.

The simulations have been run on two different clusters: (i) the Intelligent Systems Group (ISG) cluster and (ii) the Hyperion supercomputer of the Donostia International Physics Center (DIPC). In the ISG cluster, we used a node with 2 AMD EPYC 7252 multicore processors with 8 cores each and 8 NVIDIA RTX A5000 64 SM GPUs (with 128 CUDA Cores each). In Hyperion, we used two different nodes depending on whether the execution required GPUs or not. If the simulation was CPU-only, a node with 64 Intel Xeon Platinum 8362 and 2 TB of memory was used. For GPU execution, a node with 56 Intel Xeon Gold 6348, 1 TB of memory and 8 NVIDIA A100 SXM4 with 80 GB of memory each was used. In the OpenMP version, the threads are distributed on different cores; avoiding the use of SMT (simultaneous multithreading). The use of Hyperion is of special interest for two fundamental reasons: First, and due to the greater amount of computational resources, it allows us to run tests with a greater number of threads and check how far the parallelization is effective. Secondly, and unlike the runs on the ISG cluster, these nodes are not for exclusive use; that is, the simulations have coexisted with processes from other researchers and shared resources. These runs can show the behavior of the simulation in more realistic working environments.

Although we focus on execution time, we are not aiming to obtain the fastest algorithm per iteration, but rather to generate a simple benchmark to compare the future quantum version. Therefore, we are interested in understanding how execution time grows in relation to the number of iterations, which is more closely related to algorithmic complexity. For this reason, the project is compiled without applying any optimization options by the compiler.

\subsubsection{OpenMP performance}
\label{subsec:openmp_perfor}

\begin{figure} [t]
\centering
    \caption{Speed-up factor of the OpenMP version with 2 to 48 threads. These simulations have been run without swaps.}
\begin{minipage}{0.495\textwidth}
    \centering
    \subcaption{ISG cluster} \label{fig:speed_omp_isg}
    \includegraphics[width=\linewidth]{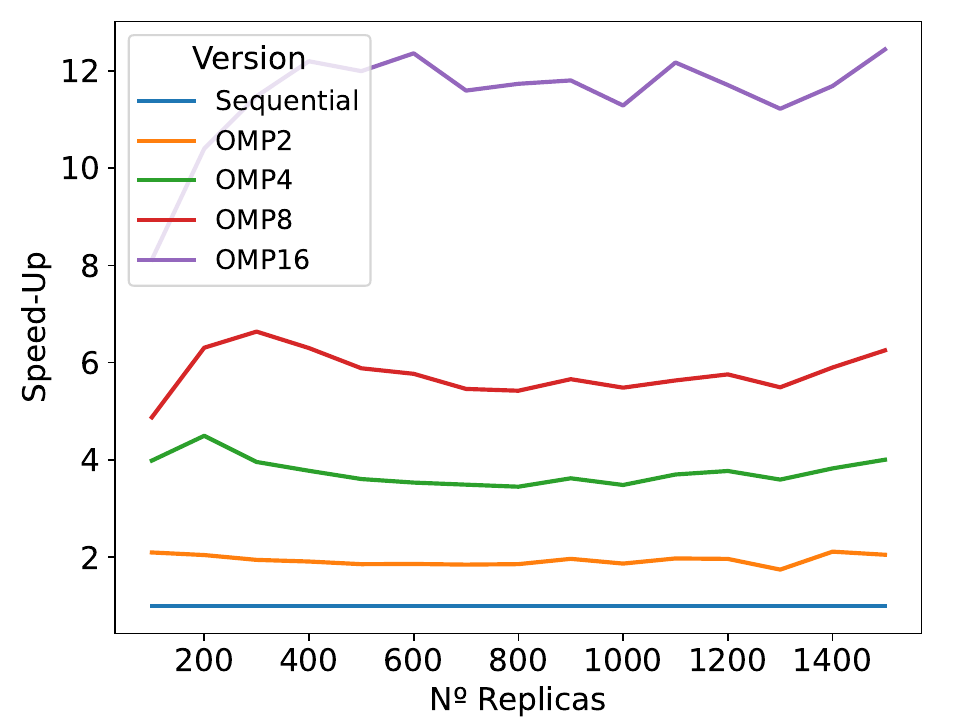}
\end{minipage}
\begin{minipage}{0.495\textwidth}
    \centering
    \subcaption{Hyperion} \label{fig:speed_omp_hyperion}
    \includegraphics[width=\linewidth]{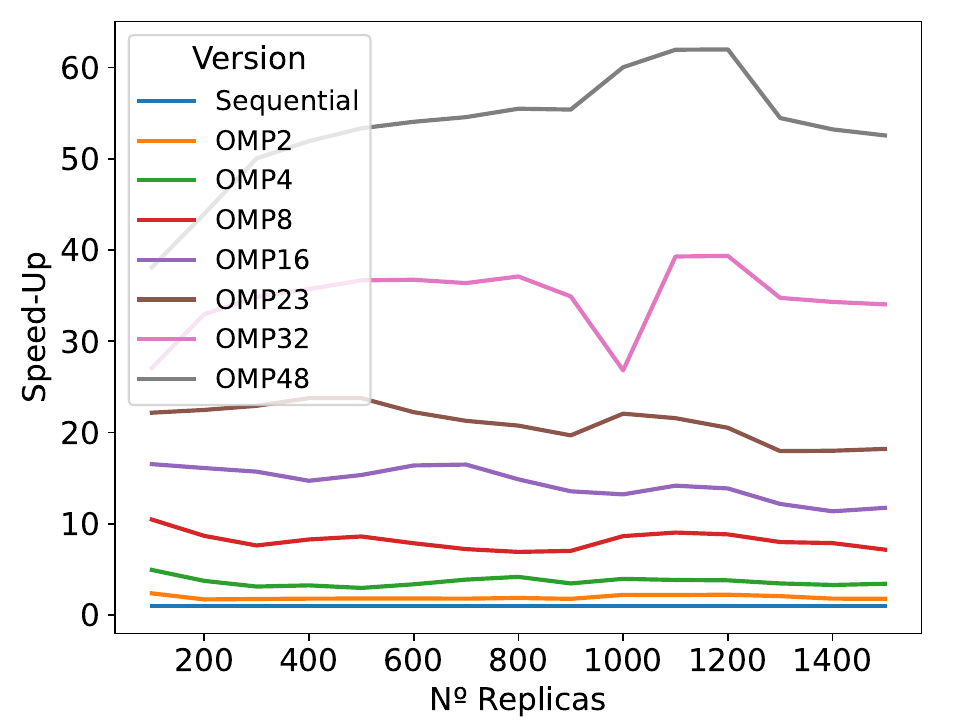}
\end{minipage}
\label{fig:speed_omp_figs}
\end{figure}

For this analysis, all the simulations were executed without swaps between replicas to remove the effects of synchronization and measure the effectiveness of the parallelization. 
Fig.~\ref{fig:speed_omp_isg} shows the speed-up factor versus the sequential version of the OpenMP versions, executed with 2, 4, 8, and 16 threads in the ISG cluster. In the first place, the addition of threads into the simulation leads to an increment in the speed-up factor; an increment independent to the number of replicas on the simulation. This behavior is interpreted as a sign of good parallelization and positive contribution of all threads to the global performance. Something to note is the reduction of the increment in the speed-up factor when the number of threads is increased. For example, executions with 2 and 4 threads obtained consistent speed-up factors of $2\times$ and $4\times$, but 8 and 16 threads obtained $6\times$ and $12\times$ speed-up factors respectively, instead of the expected $8\times$ and $16\times$.

Hyperion results, shown in Fig.~\ref{fig:speed_omp_hyperion}, have similar behavior as the ISG cluster, but with small variations. First, the effect of speed-up reduction is not present for Hyperion. The 32 and 48 threaded runs show peak speed-up factors of $38\times$ and $60\times$ respectively, much higher than the expected $32\times$ and $48\times$. Secondly, the number of replicas does seem to affect the speed-up, reducing it, but never below the expected value.

\begin{figure} [t]
\centering
    \caption{Performance of the sequential version compared to the OpenMP version with 2 to 48 threads and to the CUDA version with a block size of 32 threads.}
\begin{minipage}{0.495\textwidth}
    \centering
    \subcaption{ISG cluster} \label{fig:perfor_total_isg}
    \includegraphics[width=\linewidth]{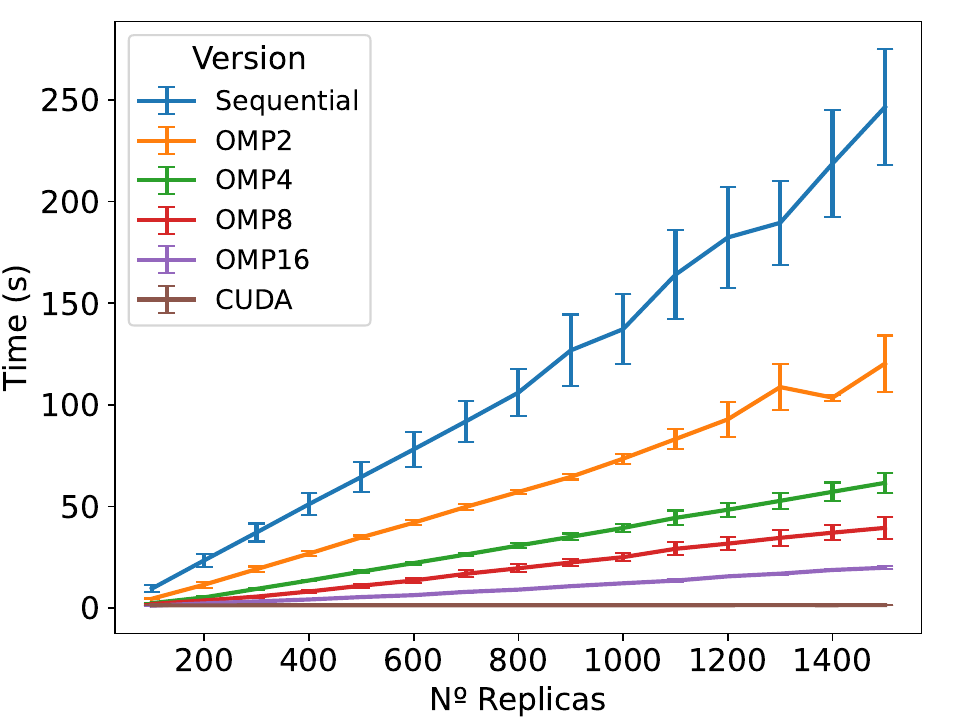}
\end{minipage}
\begin{minipage}{0.495\textwidth}
    \centering
    \subcaption{Hyperion} \label{fig:perfor_total_hyperion}
    \includegraphics[width=\linewidth]{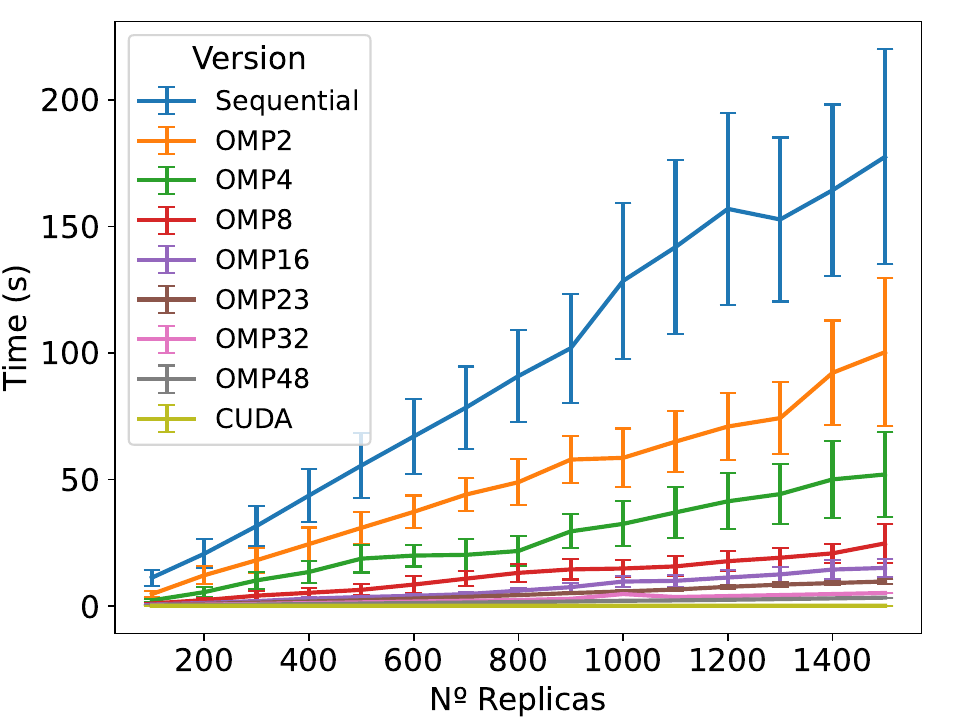}
\end{minipage}
\label{fig:perfor_total_figs}
\end{figure}

%\caption{The best result is obtained with the OpenMP version with 16 threads, with an acceleration factor of up to 12.43x in the simulation with 1500 replicas.}

%\caption{The best result is obtained with the OpenMP version with 48 threads, with an acceleration factor of up to 52.57x in the simulation with 1500 replicas.}

%%%%%

Another thing to note is the behavior of the variation of execution times between the same executions on the ISG cluster in Fig.~\ref{fig:perfor_total_isg}. It can be observed how the increase in simulated replicas makes the time between executions vary more, reaching a maximum variation of 60 seconds in the case of sequential version with 1500 replicas. In contrast, the increase in threads reduces this variation almost to 0 with simulations run with 16 and above. In general, an increase in replicas makes the executions more unstable in terms of execution time and, in contrast, the number of threads makes them more stable.

The executions time in Hyperion, showed in Fig.~\ref{fig:perfor_total_hyperion}, has roughly the same behavior as the ISG cluster, except for two aspects. First, it achieves a time reduction of up to 60 seconds, due to the higher computational power. And secondly, more noisy execution times due to the interactions with other processes coexisting in the node, all using the same resources and increasing latencies.
\subsubsection{Block size in CUDA}
\label{subsec:block_size}

\begin{figure} [t]
\centering
    \caption{Performance comparison of the CUDA version using different block sizes.}
\begin{minipage}{0.495\textwidth}
    \centering
    \subcaption{ISG cluster} \label{fig:blocksize_CUDA_bender}
    \includegraphics[width=\linewidth]{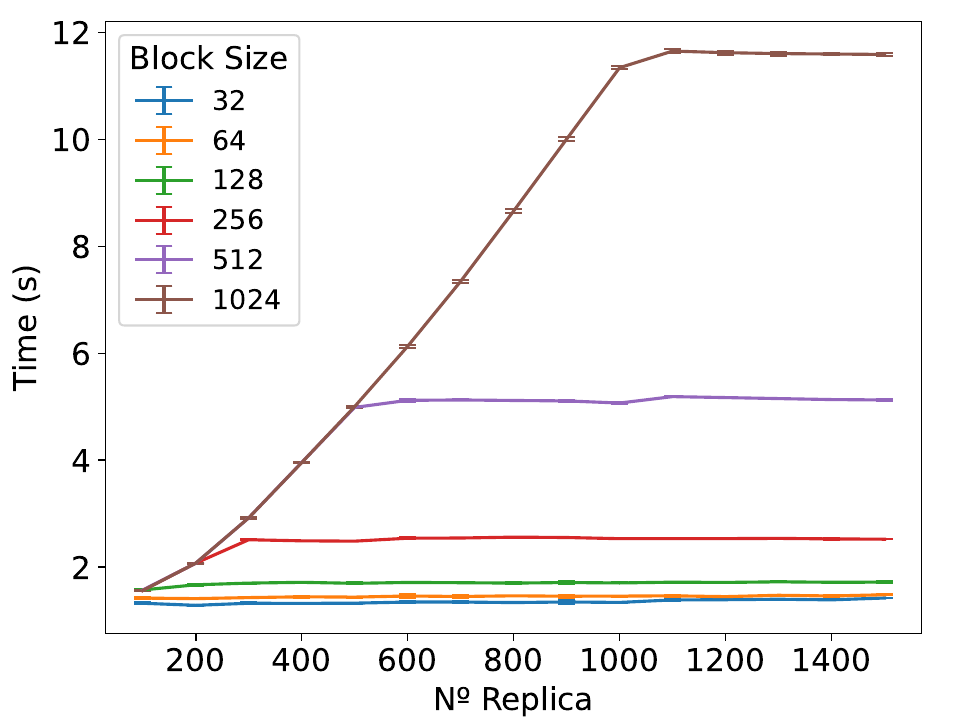}
\end{minipage}
\begin{minipage}{0.495\textwidth}
    \centering
    \subcaption{Hyperion} \label{fig:blocksize_CUDA_hyper}
    \includegraphics[width=\linewidth]{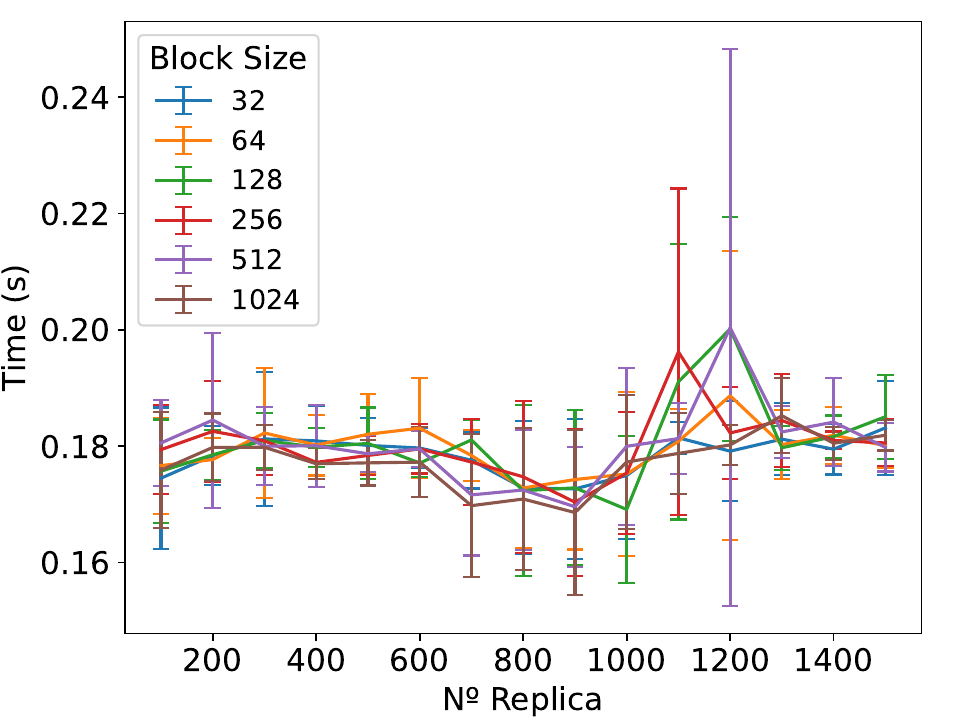}
\end{minipage}
\label{fig:perfor_total_figs}
\end{figure}

In CUDA applications the block size is crucial to obtain the maximum performance in these devices. For each application, the optimum block size may vary due to the executed {\it kernel}. For this reason, it is necessary to investigate different block sizes to get the best configuration. Fig.~\ref{fig:blocksize_CUDA_bender} shows the effect that block size has on performance when our application is run in the ISG cluster. In particular, we found out that in this set-up the optimal block size is about 32 threads, a {\it warp}. Taking into account that the GPU used for these experiments has 64 SM and each of them can run up to 128 threads simultaneously, the device can execute {\bf 8192 threads} at once. With 32 threads per block configuration, we can execute {\bf up to 2048 replicas} concurrently inside the device. The higher the block size is, the lower the occupation of the SM in the device and, in consequence, the worse results are obtained. In Hyperion  the size of the block is irrelevant for the execution time as means are very similar and the ranges of the variances overlap, as shown in Fig.~\ref{fig:blocksize_CUDA_hyper}. This can be explained by the fact that NVIDIA A100 SXM4 have 108 SM (remember that on the ISG cluster the NVIDIA RTX 5000 has 64 SM). They also have larger and faster cache memories, accelerating block switching within SM and speeding up the execution of all threads.

We can observe in Fig.~\ref{fig:perfor_total_isg} (ISG cluster) and Fig.~\ref{fig:perfor_total_hyperion} (Hyperion) that {\bf regardless} of the number of replicas of the simulation, the CUDA version maintains a stable time below two seconds in ISG cluster and below one second in Hyperion. It is expected that as the number of replicas increases, the execution time of this version will increase due to the {\bf lack of resources} to execute all replicas at the same time, making its behavior more similar to the OpenMP version. In summary, in the ISG cluster the OpenMP version is up to 12.43 times faster than the sequential version, and the CUDA version, in turn, is 13.93 times faster than the OpenMP version, i.e., {\bf 173.22 times faster} than the sequential version. In Hyperion, the OpenMP version is 52.57 times faster than the sequential version, and the CUDA version presents an execution time {\bf 986.38 times faster} than the sequential version.

\subsubsection{Impact of swaps on performance}
\label{subsec:swaps}

The results above pertain to MH/PT runs without replica swapping, i.e., each MH process acts independently. However, as previously mentioned, the goal of the PT technique is to use replica swapping to favor the exploration of the target distribution. These exchanges require synchronization between threads, which, in parallel programming, can be very costly and become the bottleneck. For this reason, it is necessary to measure the effect of swapping on the performance of the different versions. For this purpose, and as explained in the experimental setup, tests have been run with different sizes of intervals between swaps.

\begin{figure} [t]
\centering
    \caption{Mean times and standard deviations of simulation runs with 1500 replicas and varying interval sizes between exchanges. }
\begin{minipage}{0.495\textwidth}
    \centering
    \subcaption{ISG cluster} \label{fig:swaps_isg}
    \includegraphics[width=\linewidth]{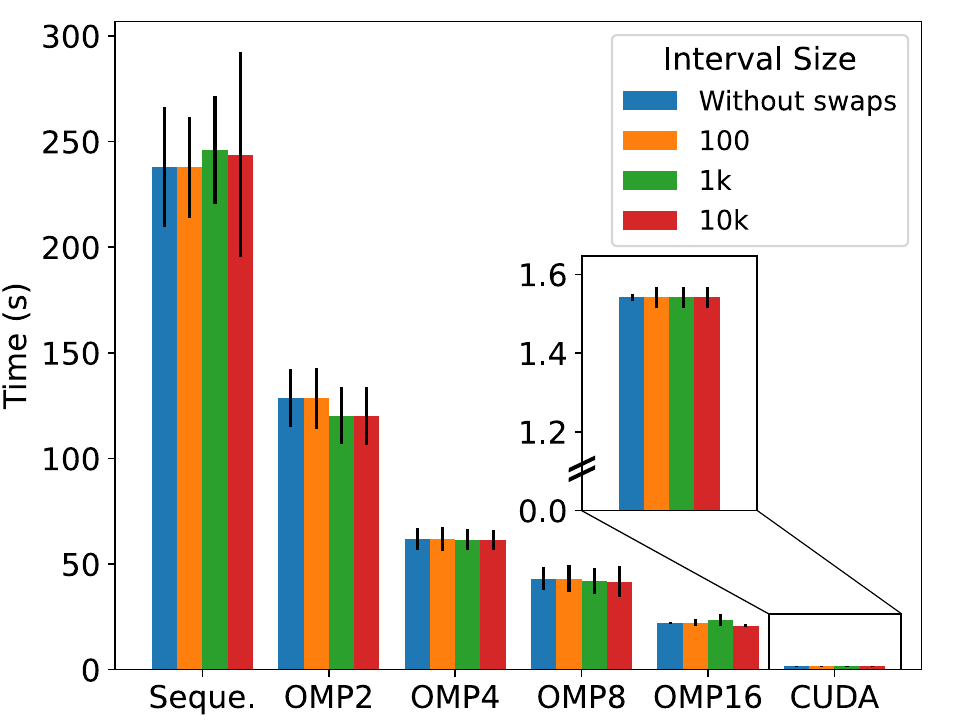}
\end{minipage}
\begin{minipage}{0.495\textwidth}
    \centering
    \subcaption{Hyperion} \label{fig:swaps_hyperion}
    \includegraphics[width=\linewidth]{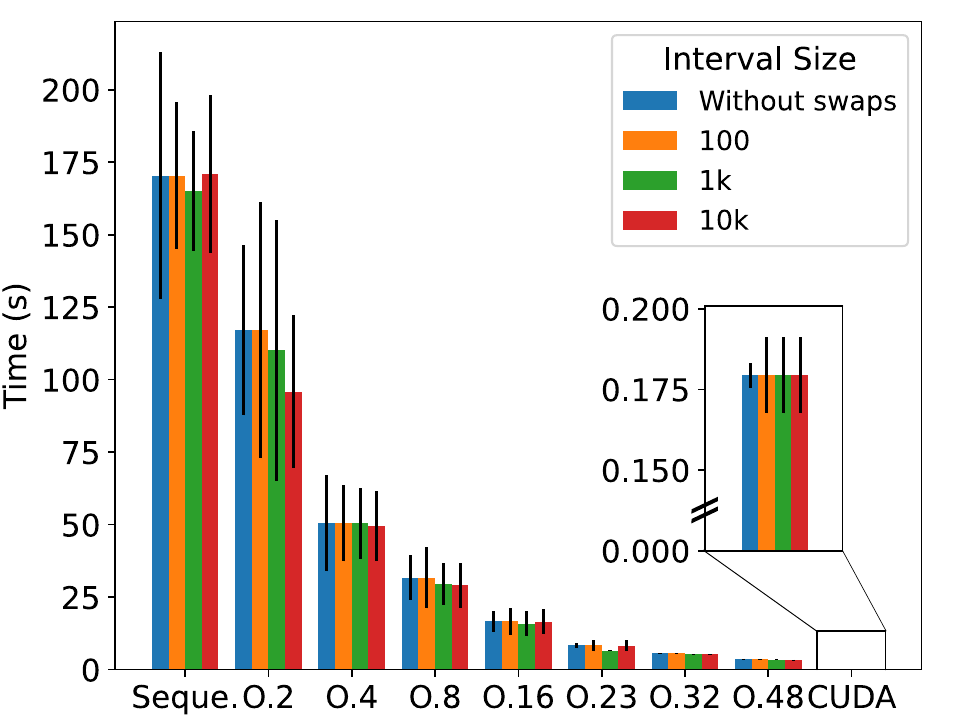}
\end{minipage}
\label{fig:swaps_figs}
\end{figure}

In Fig.~\ref{fig:swaps_isg} and Fig.~\ref{fig:swaps_hyperion} we show the execution times for different swap interval sizes for simulations with 1500 replicas in the ISG cluster and Hyperion respectively. It can be seen that regardless of the size of the interval between swaps and in any version (OpenMP, CUDA or sequential), the execution time is slightly affected, being similar in execution time to the simulation without swaps. This is due to the low ratio of swaps accepted in the simulations. The Ising model is known to be a very ''{\it glassy}'' system; neighboring replicas tend to diverge quickly, which greatly reduces the probability of accepting the swap between two replicas. It can also be observed, and following the results presented previously, that increasing the number of threads available in the simulation reduces the execution time considerably.

\section{Conclusion}
\label{sec:conclusion}

In this paper we have presented two parallel implementations of the Metropolis-Hastings (MH) algorithm using Parallel Tempering (PT) for Ising model sampling for two architectures, multicore and GPU, using OpenMP and CUDA respectively. The results of our experimental evaluation show that both implementations efficiently exploit computational resources, taking advantage of the parallel capabilities of both architectures. In both cases, considerable speedups are obtained with respect to the sequential version, reaching a maximum speedup factor of up to $52.57\times$ in the OpenMP version (with 48 cores), and up to $986.38\times$ in the CUDA version; both using the nodes of Hyperion cluster. It is also worth noting that in the GPU version, all the computation is done inside the device avoiding transfers with the {\it host}, a difference with respect to previous works. Furthermore, along with the analysis of the convergence speed of the algorithm in the Ising model, it constitutes a basic benchmark to compare the efficiency of the future quantum version of the method.

The main future direction is developing the quantum version, but there are improvements to explore in the current classical version: (i) optimization of memory usage and (ii) implementation of more complex models. Regarding the first, our implementation is very demanding in terms of memory usage, as it stores all the chains. Developing techniques that allow for better reduction or management of memory usage could alleviate this limitation, enabling the execution of larger models. As for the second, although the current implementation allows for 'inserting' and running another model, this functionality is still in a very primitive phase. Improving this aspect would expand the benchmark variety and facilitate the comparison of the efficiency of the future quantum version.

%%%%%%%%%%%%%%%%%%%%%%%%%%%%%%%%%%%%%%%%%
\begin{credits}
\subsubsection{\ackname} 
\noindent  This research is supported by the Basque Government through projects IT1504-22, KK-2023/00012, KK-2023/00090, KK-2024/00030 and KK-2024/ 00068. 
It is also supported by the grant CNS2023-144315 funded by MICIU/AEI/10. 13039/9/501100011033 and by ``European Union NextGenerationEU/PRTR''.
It is also supported by the grant PID2023-152390NB-I00 funded by the MICIU/AEI/10.13039/ 501100011033 and by ``FEDER funds''.
Dr. Javier Navaridas holds a Ramon y Cajal fellowship funded by the Spanish Ministry of Science, Innovation and Universities (RYC2018-024829-I), funded by MCIN/AEI/10.13039/501100011033 and, as appropriate, by ``ESF Investing in your future'' or by ``EU NextGenerationEU/PRTR''. 
Authors want to thank the DIPC computing centre for providing the technical and computing resources needed to carry out some of the experiments of this paper.
\subsubsection{\discintname}
The authors have no competing interests to declare that are relevant to the content of this article.
\end{credits}

%%%%%%%%%%%%%%%%%%%%%%%%%%%%%%%%%%%%%%%%%

\end{document}